\documentclass[prb,superscriptaddress,amsfonts,twocolumn,showpacs,floatfix]{revtex4}

\usepackage{graphicx}
\usepackage{dcolumn}
\usepackage{bm}
\usepackage{mathrsfs}
\usepackage{exscale}
\usepackage{color}
\begin{document}

\preprint{APS/123-QED}

\title{Low-lying excitations at the rare-earth site due to the rattling motion \\in the filled skutterudite LaOs$_4$Sb$_{12}$ 
revealed by $^{139}$La NMR and $^{121/123}$Sb NQR}

\author{Yusuke Nakai}
\altaffiliation{nakai@scphys.kyoto-u.ac.jp}
\author{Kenji Ishida}
\affiliation{Department of Physics, Graduate School of Science, Kyoto University, Kyoto 606-8502, Japan,}
\author{Hitoshi Sugawara}
\affiliation{Faculty of Integrated Arts and Sciences, University of Tokushima, Tokushima 770-8502, Japan,}
\author{Daisuke Kikuchi}
\author{Hideyuki Sato}
\affiliation{Graduate School of Science, Tokyo Metropolitan University, Hachioji, Tokyo 192-0397, Japan}

\date{\today}

\begin{abstract}
We report experimental results of nuclear magnetic resonance (NMR) at the La site and nuclear quadrupole resonance (NQR) at the Sb site in the filled skutterudite LaOs$_4$Sb$_{12}$. We found that the nuclear spin-lattice relaxation rate divided by temperature $1/T_1T$ at the La site exhibits a different temperature dependence from
that at the Sb site. Although $1/T_1T$ at the Sb site is explained by the Korringa mechanism, $1/T_1T$ at the La site
exhibits a broad maximum around 50 K, showing the presence of an additional contribution at the La site. The additional low-lying excitations observed at the La site can be understood with the relaxation from anharmonic phonons due to the rattling motion of the La atoms.
\end{abstract}

\pacs{71.27.+a, 74.25.Kc, 76.60.-k, 76.60.Gv}
\maketitle

The filled skutterudite compound $RT_4X_{12}$ ($R$=rare earth;
$T$=Fe,Ru,Os; $X$=P,As,Sb) has a unique structure in which
the $R$ ion is located in an oversized cage made of 12 $X$ atoms.
The $R$ ion is loosely bound inside of the $X_{12}$ cage because the
cage size is much larger than the $R$ ion. A weak atomic
interaction between the $R$ ion and the cage atom is considered
to cause the vibration of the $R$ ion to be highly anharmonic.
Due to this cage structure, the anharmonic vibration
with large amplitude of the $R$ ion in the cage occurs, which is
called grattling.h From a practical point of view, the rattling
attracts much attention because it may reduce thermal conductivity,
which is a key to producing high-performance
thermoelectric materials.\cite{Sales} The low thermal conductivity is
believed to result from localized vibrational modes at low
energy, attributed to the rattling.\cite{Keppens} Specific heat results suggest
a nearly dispersionless low-energy optical mode that is
characterized by an Einstein phonon mode.\cite{Takabatake} A large Debye-
Waller factor obtained with neutron diffraction experiments
indicates a large thermal displacement of the $R$ ion.\cite{Kaneko} In Raman
spectra, a second-order phonon peak appears originating
from an almost dispersionless optical mode at low energy
relevant to the rattling.\cite{OgitaJMMM,OgitaPhysica} An anomalous Debye-type dispersion
of elastic constants found in ultrasound measurements is
also attributable to the rattling.\cite{Goto}

Recently, it has been recognized that the rattling motion
might be a key phenomenon to understand the exotic heavyfermion
(HF) state in $R$Os$_4$Sb$_{12}$ ($R$=Pr, Sm, Nd). The exotic
HF state shows very different behavior from magnetic
Kondo materials such as Ce- or U-based HF compounds.
They do not exhibit $\!-\!\log T$ behavior in the resistivity.\cite{Bauer, Sanada, SugawaraNdOs4Sb12, Ho}
Furthermore, a large electronic specific-heat coefficient
[ $\gamma$=0.82 J / (K$^2$ mol)] in SmOs$_4$Sb$_{12}$ is insensitive to magnetic
fields.\cite{Sanada} These results suggest that the exotic HF state
results from a nonmagnetic origin. This contrasts with the
conventional HF state formed by magnetic interactions between
the conduction electron and localized $f$ electron ($c$-$f$
interaction). In this situation, the rattling motion of the $R$ ion
is invoked to be related to the formation of the exotic HF
state. In fact, HF states originating from anharmonic ion vibrations
have been analyzed theoretically.\cite{Hattori,Mitsumoto}

Characterizing the rattling motion is quite important to
understand the low thermal conductivity and the exotic HF
state. However, the characterization is difficult from bulk
measurements because the rattling is localized vibrational
modes at the rare-earth site. In general, one can extract the
localized vibrational modes through comparing between a
filled skutterudite and an unfilled one. However, such a comparison
is impossible when an unfilled skutterudite cannot be
synthesized or is not appropriate when filling $R$ ions change
the electronic state significantly. For these reasons, it is desirable
to compare physical properties at the $R$ site with those
at the Sb site to characterize the rattling in the same compound.
For this purpose, nuclear magnetic resonance (NMR)
and nuclear quadrupole resonance (NQR) are suitable
probes, because NMR and NQR can derive low-lying excitations
site-selectively due to the rattling motion. Although it
is desirable to perform Pr or Sm NMR measurements, these
nuclei are too magnetic to allow NMR measurements. For
this reason, we instead performed $^{139}$La NMR and $^{121/123}$Sb
NQR experiments on LaOs$_4$Sb$_{12}$ because the characteristic
motion of $R$ ions is believed to be similar for the $R$Os$_4$Sb$_{12}$ 
($R$= La, Pr, Sm, Nd) family. Because the $^{139}$La nucleus 
($I$=7/2) has an electric quadrupolar moment $Q$, the nuclear
spin-lattice relaxation rate $1/T_1$ at the La site can directly
probe the unusual dynamics of the La ions through quadrupolar
coupling of the La nucleus to the electric field gradient 
(EFG).

In this paper, we report results of $^{139}$La NMR and 
$^{121/123}$ Sb NQR measurements on LaOs$_4$Sb$_{12}$. We found that
$1/T_1T$ at the La site exhibits a different temperature dependence
from that at the Sb site. By subtracting the conduction
electron contribution from the observed $1/T_1T$ at the La site,
we obtained additional relaxation at the La site $^{139}(1/T_1T)_{\rm add}$.
$^{139}(1/T_1T)_{\rm add}$ is nearly constant at high temperatures and decreases
rapidly below about 50 K. We show that this temperature
dependence can be explained quantitatively by a
recent theoretical calculation based on the Raman process of
anharmonic phonons due to the rattling motion.\cite{Dahm} From the
present result, we suggest that the low-lying excitations
caused by the rattling of $R$ ions might be important to understand the exotic HF state in $R$Os$_4$Sb$_{12}$ ($R$= Pr, Sm, Nd).

Single crystals of LaOs$_4$Sb$_{12}$ were grown with the Sb-flux 
method\cite{Sugawara-de-Haas} and were powdered for NMR and NQR measurements.
The observation of de Haas-van Alphen oscillations
demonstrates that we were using a high-quality sample.\cite{Sugawara-de-Haas}
NMR and NQR measurements were performed using a conventional
pulsed spectrometer in the temperature range of
$T$= 0.6 $!-!$ 250 K using a $^3$He-$^4$He-dilution refrigerator. The
nuclear spin-lattice relaxation rate $1/T_1$ of La nuclei was
measured using the saturation recovery method and was
uniquely determined with a single component. $1/T_1$ of $^{121}$Sb($^{123}$Sb) NQR was measured around 84 (78) MHz using a
2$\nu_Q$ (3$\nu_Q$) transition.\cite{Yogi} The magnetic susceptibility $\chi$ was
measured for the powdered sample with a superconducting
quantum interference device (SQUID) magnetometer (Quantum
Design MPMS).

Figure 1 shows the temperature dependence of $1/T_1T$ at
the La site under various magnetic fields, along with that at
the $^{121}$Sb site (open circles). $^{121}(1/T_1T)$ is consistent with the
previous report.\cite{Kotegawa} We found that $1/T_1T$ at the La and Sb sites
exhibits different $T$ dependences; the broad maximum
around 50 K appears only at the La site. This is a marked
contrast to the La-based skutterudite LaFe$_4$P$_{12}$, in which
$1/T_1$ at both La and P sites exhibits the same temperature
dependence.\cite{Nakai, NakaiRattling}
\begin{figure}
\begin{center}
\includegraphics[width=8.8cm]{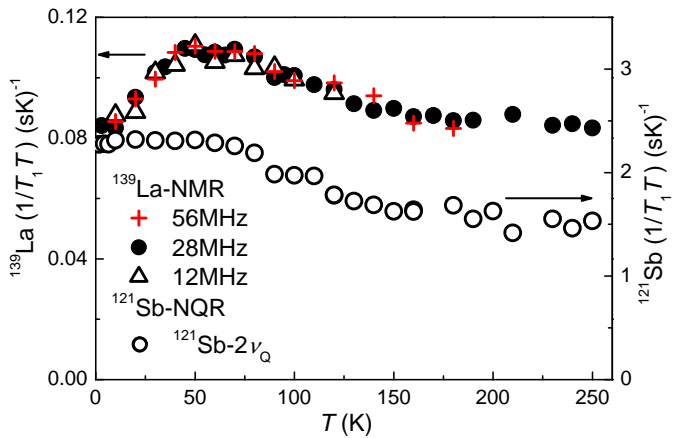}
\end{center}
\caption{(Color online) Temperature dependence of $1/T_1T$ at the La and Sb site under various magnetic fields for La NMR. 
The broad maximum is found around 50 K only at the La site.}
\end{figure}

To clarify the origin of the site-dependent $1/T_1T$, we measured
the bulk susceptibility $\chi$ and the Knight shift at the La
site ($^{139}K$) as shown in Fig.~2. From a good linear relation
between $^{139}K$ and $\chi$, the temperature dependence of $\chi$ is not
affected by impurity phases, but is intrinsic behavior of
LaOs$_4$Sb$_{12}$. It is shown that $1/T_1T$ at the Sb site exhibits a
temperature dependence similar to $\chi^2$, but $1/T_1T$ at the La
site does not. In fact, $^{121}(1/T_1T)$ is scaled with $\chi^2$ as shown
in Fig.~3. Furthermore, the $^{121/123}$Sb-isotopic ratio of $1/T_1$ in
the whole temperature range is consistent with the case of
magnetic relaxations (=$(^{121}\gamma/^{123}\gamma)^2$ = 3.4) as shown in the
inset of Fig.~3. These results indicate that $1/T_1T$ at the Sb site
is governed by magnetic fluctuations, which are related to
the static susceptibility through the Korringa mechanism.
Note that the temperature dependence of $1/T_1T$ at the Sb site 
is considered to originate from narrow Os-$5d$ bands, which is
suggested by band calculations.\cite{Harima} Then, it seems natural to
consider that the broad peak of $1/T_1T$ at the La site is ascribed
to additional low-lying excitations at the La site. The
field-independent $1/T_1T$ (see Fig.~1) implies that the lowlying
excitations arise from nonmagnetic origin.
\begin{figure}
\begin{center}
\includegraphics[width=8.5cm]{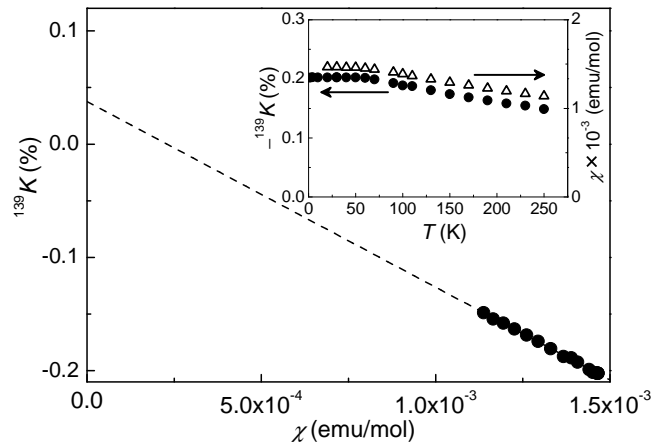}
\end{center}
\caption{\label{fig:epsart2} The $^{139}$La Knight shift $^{139}K$ is plotted against the bulk susceptibility $\chi$. The hyperfine coupling constant at the La site is estimated to be $-$10.4 kOe/$\mu_B$. 
Inset: Temperature dependence of $\chi$ and $^{139}K$.}
\end{figure}

\begin{figure}
\begin{center}
\includegraphics[width=8cm]{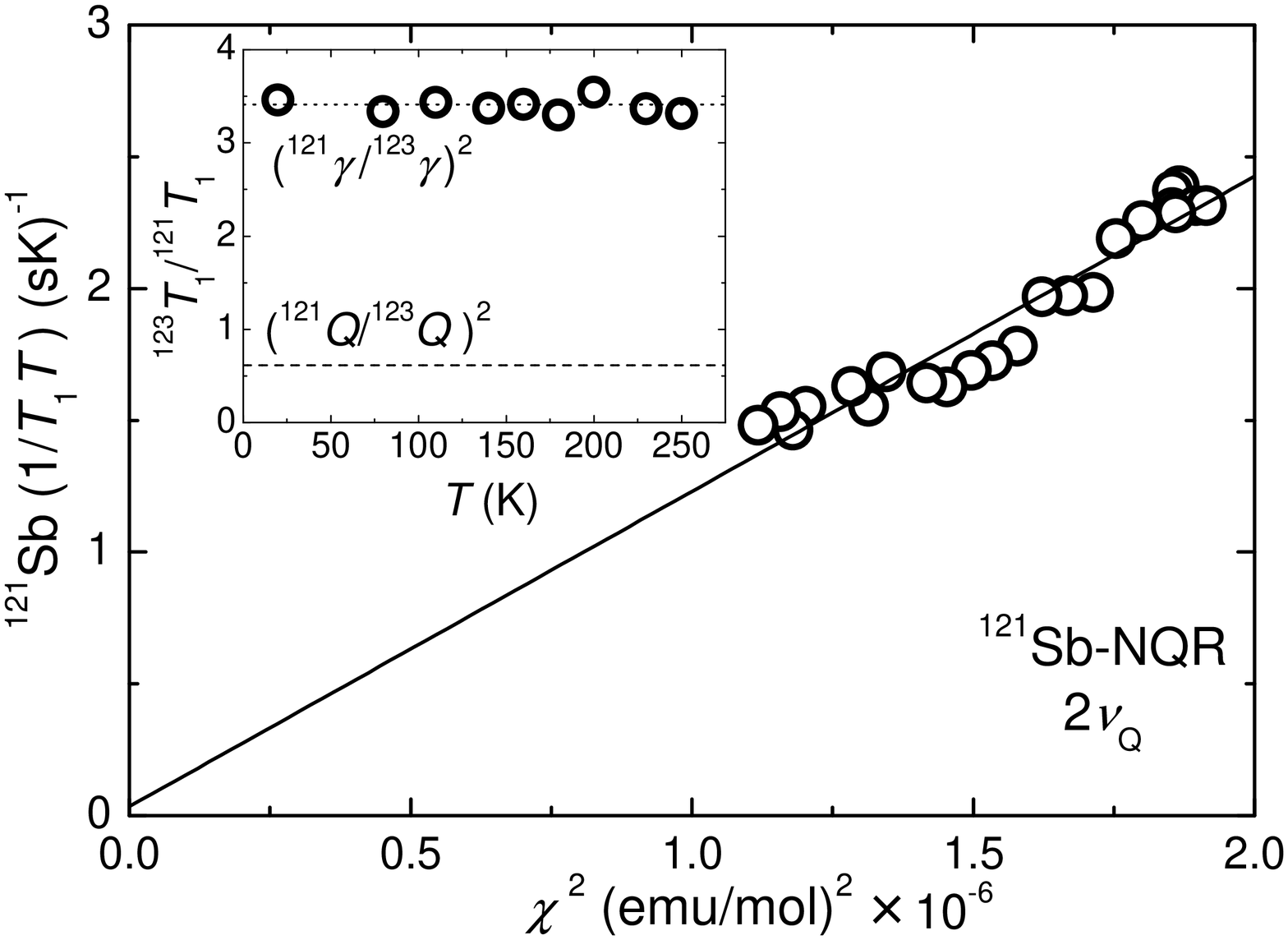}
\end{center}
\caption{\label{fig:epsart3}Plot of $1/T_1T$ of Sb against $\chi$ squared. 
The Korringa relaxation mechanism governs $1/T_1T$ at the Sb site. Inset: the isotropic ratio of $1/T_1$ for the Sb nucleus. The dotted (dashed) line shows the squared ratio of the nuclear magnetic (quadrupolar) moments. }
\end{figure}

\begin{figure}
\begin{center}
\includegraphics[width=8.2cm]{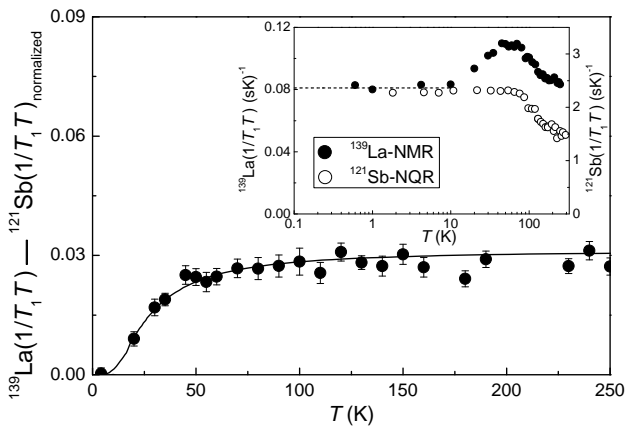}
\end{center}
\caption{The difference between $1/T_1T$ of La and normalized $1/T_1T$ of $^{121}$Sb, which is considered to arise from the rattling motion of the La ions. The solid line is the calculated result cited from Ref.~14 [$\beta=2.0$, $\omega_r(T=0)=60$ K], 
which is consistent with the experimental data. }
\end{figure}
Next, we analyze the low-lying excitations using the temperature
dependence of $1/T_1T$ at the La site. We consider
that $1/T_1T$ at the La site is governed not only by the Korringa
mechanism but also by local fluctuations of the EFG
due to the rattling motion of the La ions. In fact, ultrasonic
dispersion is found around 50 K in LaOs$_4$Sb$_{12}$,\cite{Nemoto} which is
considered to originate from the rattling motion of the La
atoms as in PrOs$_4$Sb$_{12}$.\cite{Goto} We assume that the fluctuations of
the EFG disappear at low temperatures and that the Korringa
mechanism dominates the relaxation at the La site at low
temperatures, because $1/T_1T$ at the La sites becomes constant 
below 10 K down to 0.6 K as $1/T_1T$ of $^{121}$Sb does as
shown in the inset of Fig.~4. To estimate the contribution
from the Korringa law at the La site, we normalized $1/T_1T$
of $^{121}$Sb to make it equal to that of La below 10 K, as shown
in the inset of Fig.~4. Then, we obtained the additional fluctuations
$(1/T_1T)_{\rm add}$, which is considered to arise from EFG
fluctuations related to the rattling, as shown in Fig.~4: nearly
constant at high temperatures and decreasing rapidly below
50 K. Nuclear spin relaxation due to phonons is very rare in
metals, but the small value of $(1/T_1T)_{\rm add}$ ($\sim 0.03$ s$^{-1}$K$^{-1}$)
can be explained in terms of phononic relaxation. In fact,
recent observations of NMR relaxation rates at the K site in
the $\beta$-pyrochlore superconductor KOs$_2$O$_6$ have been demonstrated
to be entirely dominated by the vibration of the K ion
via coupling of the EFG to the nuclear quadrupole moment.\cite{Yoshida}
The phononic contribution at the $^{41}$K site in KOs$_2$O$_6$ is
$1/T_1T\sim 0.009 $s$^{-1}$K$^{-1}$, which is the same order of magnitude
as $(1/T_1T)_{\rm add}$ at the La site in LaOs$_4$Sb$_{12}$. However, the EFG
fluctuations of the La atom are not the usual harmonic
phonons. If one assumes NMR relaxation from the Raman
process of harmonic phonons, the obtained temperature dependence
of $(1/T_1T)_{\rm add}$ cannot be explained: $1/T_1T$ is proportional
to $T$ at high temperatures in the Raman process for
harmonic phonons.\cite{Van} The inconsistency invokes the need to
take into account the anharmonicity that is one of the important
characteristics of the rattling motion.

Quite recently, Dahm and Ueda calculated NMR relaxation
due to coupling to a strongly anharmonic and damped
phonon mode, and explained successfully the temperature
dependence of $1/T_1$ at the K site in KOs$_2$O$_6$.\cite{Dahm} In their notation,
their Hamiltonian including anharmonic term is described
as follows:
\begin{equation}
H=\frac{p^2}{2M}+\frac{1}{2}ax^2+\frac{1}{4}bx^4,
\end{equation}
They treat this Hamiltonian in a self-consistent quasiharmonic
approximation, resulting in an effective harmonic Hamiltonian
\begin{equation}
H=\frac{p^2}{2M}+\frac{1}{2}M\omega_0^2x^2,
\end{equation}
where the effective phonon frequency is $\omega_0$. Then they obtained
a nonlinear equation of $\omega_0$:
\begin{equation}
\left(\frac{\omega_0}{\omega_{00}}\right)^2 = 1+ \beta \frac{\omega_{00}}{\omega_0}\left(\frac{1}{e^{\hbar\omega_0/k_BT}-1}+\frac{1}{2}-\frac{1}{2}\frac{\omega_0}{\omega_{00}}\right)
\end{equation}
where $\beta$ represents the magnitude of anharmonicity and $\omega_{00}=\omega_{0}(T=0)$. This equation demonstrates that the effective phonon frequency $\omega_{0}$ has a notable temperature dependence due to the anharmonicity (see Fig.~1 in Ref.~14). Due to the anharmonicity of the phonon mode, the temperature dependence of $1/T_1T$ for the Raman process is qualitatively different
from that of harmonic phonons. $1/T_1T\propto T$ at high temperature for harmonic phonons changes to $1/T_1T\sim $const at high $T$ for anharmonic phonons. They pointed out that the two-phonon Raman process dominates NMR relaxation instead of the direct process. The frequency-independent $1/T_1T$ at the La site (see Fig.~1) is consistent with the Raman process as studied in Ref. 23. The solid line in Fig.~4 is a calculation in Ref.~14 assuming $\beta=2.0$, $\omega_0(T=0)=60$ K. The calculation is consistent with the experimental data. Furthermore, the low temperature value of the obtained phonon frequency $\omega_0(T=0)=60$ K is quite consistent with a low Einstein temperature of 59 K reported in specific-heat experiments,\cite{Takabatake} a flat phonon-dispersion energy about 50 K deduced from the second-order phonon peak related to vibration of the La ions in Raman scattering experiments,\cite{OgitaPhysica, OgitaJMMM} and about 5 meV ($\sim 60$ K) by {\it ab initio} calculation.\cite{Hasegawa} In their calculation, anharmonicity (corresponds to $\beta$) plays a crucial role to account for the observed temperature dependence of
NMR relaxation. The anharmonicity in the Os$_4$Sb$_{12}$ family actually has been observed experimentally from neutron\cite{Iwasa, Kaneko2} and Raman scattering measurements.\cite{Udagawa} The value of $\beta=2.0$ in LaOs$_4$Sb$_{12}$ is smaller than that of KOs$_2$O$_{6}$($\beta=6.27$), suggesting that the anharmonicity of LaOs$_4$Sb$_{12}$ is smaller than
that of KOs$_2$O$_{6}$. This is reasonably understood by the fact that the mass of the K ion is much smaller than that of the La ion. The temperature dependence of the resistivity suggests a stronger anharmonicity in KOs$_2$O$_{6}$ than in LaOs$_4$Sb$_{12}$. A concave-downward temperature dependence is observed in the resistivity in KOs$_2$Sb$_{12}$\cite{Hiroi}, which is considered to be related to anharmonic phonons, but just a small shoulder is observed in LaOs$_4$Sb$_{12}$.\cite{Sugawara} Therefore, we consider that the additional relaxation at the La site in LaOs$_4$Sb$_{12}$ originates from the anharmonic phonons due to the rattling motion.

The localized and strongly anharmonic phonon modes of the rare-earth ions observed in LaOs$_4$Sb$_{12}$ are presumably a common feature in the Os$_4$Sb$_{12}$ cage. We consider that the rattling motion is related to the exotic HF state in $R$Os$_4$Sb$_{12}$ ($R$= Pr, Sm, and Nd). In fact, a large mass enhancement is observed in KOs$_2$O$_{6}$, in which active rattling motion is expected.\cite{bruhwiler} If the rattling is related to the formation of the heavy-electron mass, the rattling motion of $f$ electrons at rare-earth sites would be important for the exotic HF state in $R$Os$_4$Sb$_{12}$, because such exotic HF behavior is not observed
in LaOs$_4$Sb$_{12}$ without $f$ electrons.\cite{Sugawara} Very recently, NQR measurements 
surements on SmOs$_4$Sb$_{12}$ have suggested that charge fluctuations originating from the rattling motion of Sm ions might play an important role in forming the exotic HF state.\cite{KotegawaSmOs4Sb12} Elucidating the relationship between the rattling motion of the $f$ electrons and the exotic HF state is a fascinating problem in $R$Os$_4$Sb$_{12}$ ($R$=Pr, Sm, and Nd), which deserves the further experimental and theoretical study.

In conclusion, on the basis of the different temperature dependence of $1/T_1T$ at the La and Sb sites, we found that additional low-lying excitations at the La site exist in LaOs$_4$Sb$_{12}$. The temperature dependence of $1/T_1T$ at the La site is consistently understood in terms of the Raman process of anharmonic phonons, attributed to the rattling motion of the La ions. The low-lying excitations caused by the rattling are suggested to be related to the exotic heavy-fermion state
in the $R$Os$_4$Sb$_{12}$ family. Further study is required to clarify
the relationship.

We thank T. Dahm and M. Takigawa for valuable discussions.
Two of the authors (Y.N. and K.I.) thank K. Kitagawa and Y. Maeno for their experimental support. This work was supported by a Grant-in-Aid for the 21st Century COE gCenter for Diversity and Universality in Physicsh from the Ministry of Education, Culture, Sports, Science and Technology
(MEXT) of Japan, by Grants-in-Aid for Scientific Research from the Japan Society for the Promotion of Science (JSPS), and by Grants-in-Aid for Scientific Research in Priority Area gSkutteruditeh (Nos. 16037208 and 15072206).


\end{document}